\newcommand{\CLASSINPUTtoptextmargin}{0.75in}
\newcommand{\CLASSINPUTbottomtextmargin}{1in}
\documentclass[conference]{IEEEtran}
\IEEEoverridecommandlockouts
\usepackage{amsmath,amssymb,amsfonts}
\usepackage{algorithmic}
\usepackage{graphicx}
\graphicspath{{./Figures/}}
\usepackage{textcomp}
\usepackage{xcolor}
\usepackage{subfiles} 
\usepackage{subfig}
\usepackage{color,soul}
\usepackage{cite}
\usepackage{amsmath}
\usepackage{multirow}
\usepackage{fancyhdr}

\begin{document}
\bstctlcite{IEEEexample:BSTcontrol}

\setlength{\columnsep}{0.21 in}
\def\PL{\mathrm{PL}}
\def\dB{\mathrm{dB}}

\pagestyle{fancy}
\fancyhead[C]{Confidential}
\setlength{\unitlength}{1mm}
\setlength{\columnsep}{0.25in}
\fancyhf{}
\renewcommand{\headrulewidth}{0pt}
\renewcommand{\footrulewidth}{0pt}

\fancyhf{}
\renewcommand{\headrulewidth}{0pt}
\renewcommand{\footrulewidth}{0pt}

\fancypagestyle{firststyle}{
	\fancyhf{}
	\fancyhead[L]{K. F. Nieman, O. Kanhere, R. Shiu, W. Xu, C. Duan, and S. S. Ghassemzadeh, ``An Adaptive cmWave/FR3 Channel Sounder for
Integrated Sensing and Communication,'' \textit{in GLOBECOM 2025 - 2025 IEEE Global Communications Conference, Taipei, Taiwan, Dec 2025, pp.
1–6. } }    %
	\renewcommand{\footrulewidth}{0pt}
	\renewcommand{\headrulewidth}{0pt}
}

\title{An Adaptive cmWave/FR3 Channel Sounder for \\ Integrated Sensing and Communication}

\author{\IEEEauthorblockN{Karl F. Nieman\IEEEauthorrefmark{2}, Ojas Kanhere\IEEEauthorrefmark{2}, Run-Kai Shiu\IEEEauthorrefmark{1},
 Wei-Jie Xu\IEEEauthorrefmark{1}, Chien-Yu Duan\IEEEauthorrefmark{1}, and Saeed S. Ghassemzadeh\IEEEauthorrefmark{2}}
\IEEEauthorblockA{\IEEEauthorrefmark{2}AT\&T Labs, Austin, TX USA\\
\IEEEauthorrefmark{1}Auden Techno Corp. - Smart Link Business, Taoyuan City, Taiwan\\
Email: \IEEEauthorrefmark{2}\{kn2644, ok067s, sg2121\}@att.com,
\IEEEauthorrefmark{1}\{jeff.shiu, wayne.xu, jason.duan\}@auden.com.tw}}

\maketitle

\thispagestyle{firststyle}

\begin{abstract}
In this paper, we present an advanced channel sounding system designed for sensing and propagation experiments in all types of cellular deployment scenarios. The system's exceptional adaptability, high resolution, and sensitivity makes it an invaluable tool for utilization in a variety of indoor and outdoor measurement campaigns.  The sounder has a  2.5 ns delay resolution, 170 dB path loss measurement capability and is able to measure a {360\textdegree} power-angular delay profile of the channel in less than 0.9 ms.
Additionally, the system can be easily reconfigured to measure different frequency bands by changing the RF front-end antennas. This versatile sounder is suitable for double directional channel sounding, high-speed vehicular experiments such as vehicle-to-vehicle and vehicle-to-infrastructure communications, and integrated communication and sensing experiments. 
\end{abstract}


\section{Introduction}
\label{sec:intro}
Integrated sensing and communication (ISAC) is an emerging area of research focused on combining communication and sensing functionalities into a unified system. This integration is a key feature anticipated for 6G (sixth generation) wireless technology~\cite{NGA_2024}. ISAC aims to utilize existing telecommunication infrastructure for RF sensing, enabling the detection of passive targets. Many use cases have been identified including object detection, tracking, and remote environmental monitoring with applications spanning public safety, health monitoring, defense, etc.~\cite{3gpp.22.837}. Frequency range 3 (FR3), which currently spans 7-24 GHz, is being considered for ISAC due to the wide bandwidths and improved propagation characteristics versus millimeter wave bands~\cite{NGA_2024}.

The 3GPP Spatial Channel Model (SCM) is a widely adopted wireless channel model for radio access networks (RAN). The SCM accounts for large-scale and small-scale fading, delay spread, and other critical channel characteristics essential for RAN system design. Initially developed for FR1 frequencies (up to 6 GHz), the SCM was later extended to cover FR2 frequencies (24.25-52.6 GHz) and frequencies up to 100 GHz~\cite{3gpp.38.901}. Although the SCM is well-established for FR1 and FR2, significant adjustments may be required to make it suitable for the FR3 spectrum, particularly for ISAC applications. In response, many institutions had begun characterizing wireless channels in FR3 bands~\cite{Abbasi_2024,Shakya_2024_01,Roivainen_2017,Miao_2023} as well as specific ISAC uses cases~\cite{NGA_2024}. 3GPP has ongoing studies on ISAC as well as extending SCM to FR3 as part of 5G-Advanced and 6G.

\begin{table}[b]
	\caption{FR3 channel sounders.}
	\begin{center}
		\begin{tabular}{| c | c | c | c | c | c |}
			\hline
			    \multirow[c]{3}{0.4in}{\textbf{Channel Sounder}} & \textbf{3dB}    &  \multicolumn{2}{|c|}{\textbf{field-of-view}}&     \textbf{deploy-}                       & \multirow[c]{3}{0.4in}{\centering \textbf{single~3D scan~time$^2$}}\\ 
&\textbf{bandwidth}& \multicolumn{2}{|c|}{\textbf{at RX (\textdegree)}}        & \textbf{ment}      & \\\cline{3-4}
& \textbf{(MHz)}     & \textbf{az} & \textbf{el} & \textbf{scenarios$^1$}      & \\\hline
            USC~\cite{Abbasi_2024} & 8000 & $\pm$180 & $\pm$20 & B & minutes\\
            NYU~\cite{Shakya_2024_01,Shakya_2024_outdoor} & 500 & $\pm$180 & $\pm$30 & AB & minutes \\
            Oulu~\cite{Roivainen_2017} & 500 & $\pm$180 & $\pm$45 & AB & minutes \\
            BUPT~\cite{Miao_2023} & 500 & $\pm$180 & $\pm$32 & AB & $\dagger$ \\
            This work & 400 & $\pm$180 & $\pm$30 & ABCDE & 0.5-0.9 ms \\
						\hline
			\hline
            \multicolumn{6}{l}{\scriptsize$^1$Deployment scenarios: A - indoor, B - outdoor microcell, C - outdoor macrocell,} \vspace{-0.03in}\\
            \multicolumn{6}{l}{\scriptsize \hspace{0.055in}D - mobile $\ge$ 10 km/h, E - monostatic and bistatic sensing $^2$$\dagger$ non-directional} \\
		\end{tabular}
		\label{tab:sounders}
	\end{center}
    \vspace{-0.11in}
\end{table}

Recent work on FR3 focuses on communication channels~\cite{Abbasi_2024,Shakya_2024_01,Shakya_2024_outdoor,Roivainen_2017,Miao_2023} with some work on ISAC channels~\cite{NGA_2024}. Existing FR3 sounders are based on vector network analyzers (VNAs)~\cite{Abbasi_2024,Roivainen_2017,Miao_2023} or sliding correlator~\cite{Shakya_2024_01,Shakya_2024_outdoor} with mechanical steering. The mechanically steered antennas require measurement times on the order of minutes which make mobile measurements (a desirable feature for ISAC) untenable. In ~\cite{Caudill_2021}, it is shown that electronically steered antennas can increase measurement speed but no measurements are currently available at FR3 frequencies. In addition, limited dynamic range or other factors (e.g. tethering for VNA cable) limits use cases and operating range. Work in \cite{Abbasi_2024} utilized a bandwidth of 8000 MHz which makes coordination with incumbent license holders challenging. A summary of the capabilities of prior channel sounders is provided in Table~\ref{tab:sounders}. 

To overcome these limitations, we have developed an advanced channel sounder that is adaptable to a wide range of deployment scenarios, frequencies, and experiments. Our sounder is implemented using phased array antennas, supports duplex operation, and is capable of scaling to multiple transmitter (TX) and/or receiver (RX) nodes over wide areas, while working within FCC restrictions. 

The remainder of this paper is organized as follows: Section \ref{sec:implementation} describes the design and implementation of the channel sounder; Section \ref{sec:cal} covers the system calibration and verification procedures; Section \ref{sec:field_test} presents the principal performance results; and Section \ref{sec:conclusion} provides concluding remarks and outlines directions for future research.

\section{The Channel Sounder Implementation}
\label{sec:implementation}

A channel sounder needs to be designed with the right hardware and software to meet the requirements of the experiments it will be used for. The channel sounder presented in this paper is designed to characterize locations where service providers may deploy wireless networks, such as indoor hotspot, outdoor microcell, outdoor macrocell environments, high-speed mobility such as Vehicle-to-Vehicle (V2V) and Vehicle-to-Infrastructure (V2I) environments, as well as ISAC-specific configurations (e.g. base station monostatic sensing). To address these use cases, our channel sounder was designed to have a high path loss measurement capability, the ability to measure omnidirectional (omni) angle of arrival/departure (AOA/AOD) and be able to measure the fast channel variations over short time periods or distances, while meeting the FCC spectral power transmission restrictions~\cite{fcc_license}. 

\begin{figure}[t]
\centering
\subfloat[Indoor channel sounder mounted on mobile carts]{%
  \includegraphics[clip,width=0.37\textwidth]{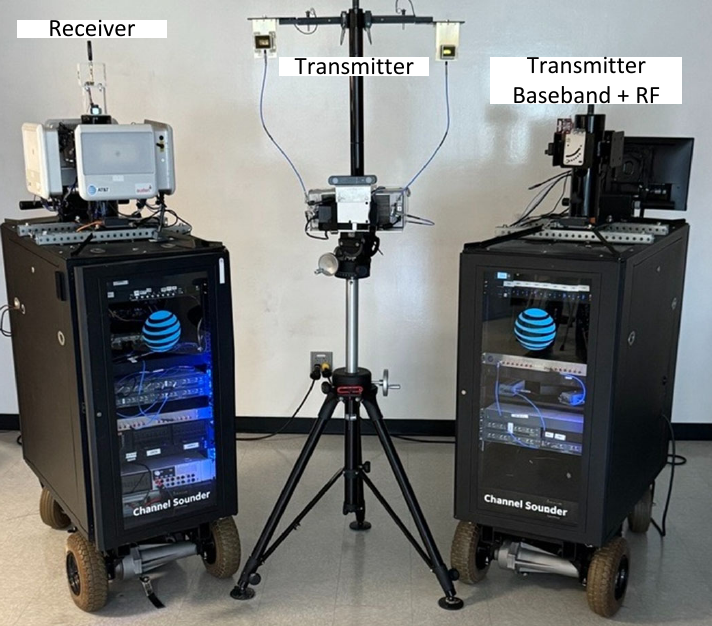}%
  \label{fig:CS-Indoor}
}
\hfill%
\subfloat[Outdoor channel sounder mounted on vehicles]{%
    \includegraphics[clip,width=0.37\textwidth]{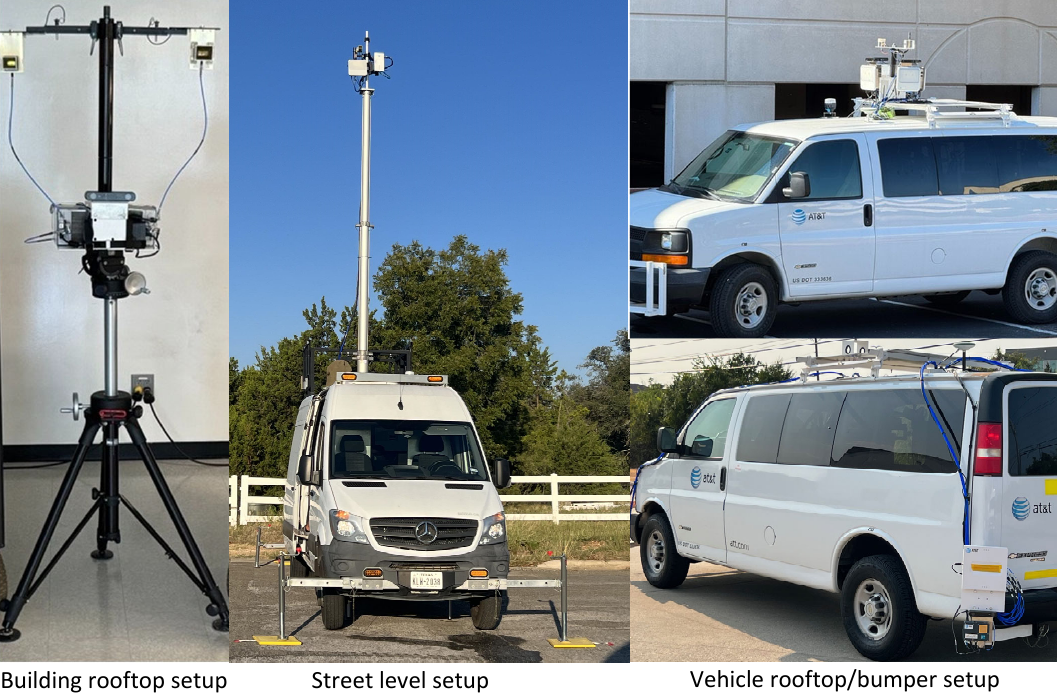}%
    \label{fig:CS-Outdoor}

}

\caption{The reconfigurable channel sounder can operate indoor and outdoors}
\end{figure}

The transceiver for our indoor sounder, shown in Fig. \ref{fig:CS-Indoor}, is assembled on two mobile carts, equipped with an electric motor for ease of transport. The transmit antenna platform, designed for simultaneous dual frequency transmission, is mounted on a tripod at a height of 3 m. The receiving arrays and omni antenna platforms are placed on a similar cart with arrays in a four 90$^{\circ}$ sector configuration. 

\begin{figure}[t]
\centering
  \includegraphics[clip,width=0.25\textwidth]{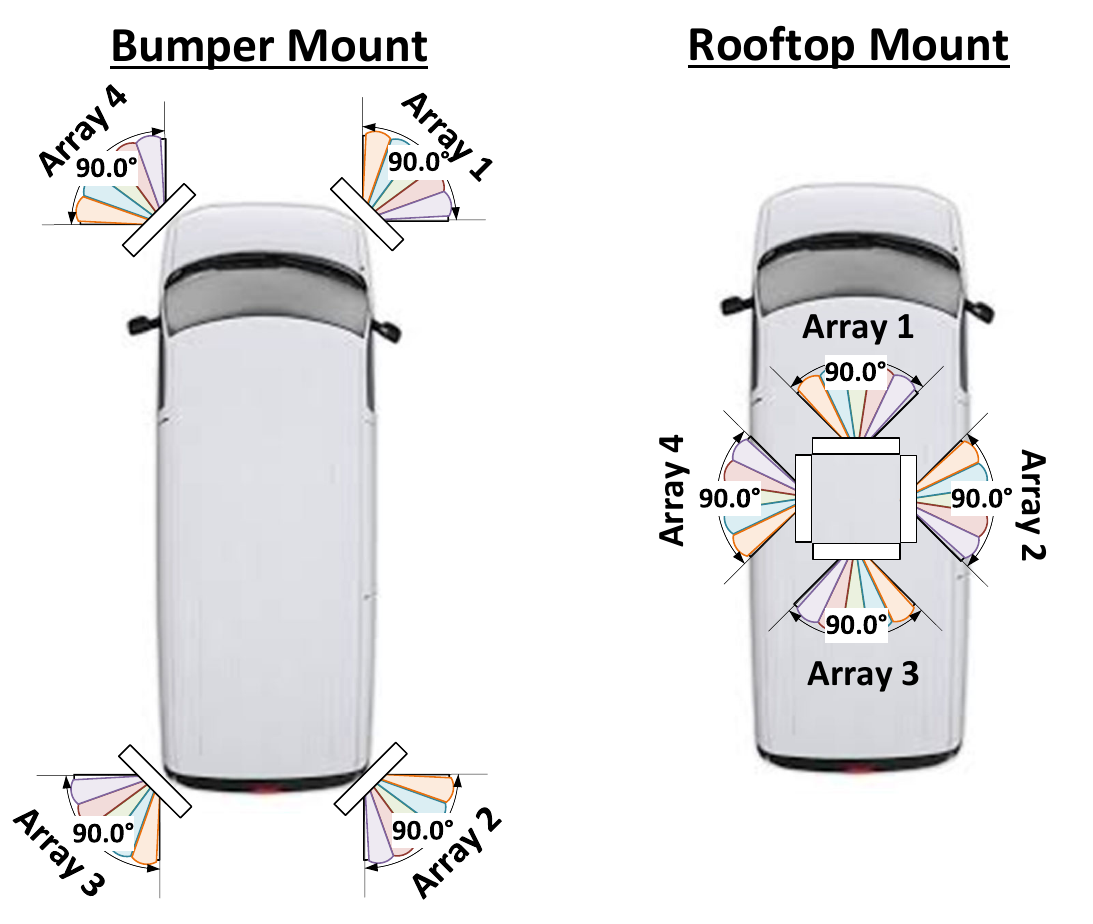}%
  \caption{Outdoor phase array orientation}
    \label{fig:PAA-platform}

\end{figure}

The outdoor platform uses the same array configuration as the indoor platform. Depending on the needs of the experiment, the receiver arrays can be mounted on the bumper or on the rooftop at a height of 0.5 m or 2 m, as seen in Fig. \ref{fig:CS-Outdoor} and Fig.~\ref{fig:PAA-platform}. The transmit antenna heights can be mounted on a tripod (for roof top deployment) for heights 10-35 m or a pneumatic mast, with typical heights of 4-10 m. 

The channel sounder employs a superheterodyne architecture, which consists of a baseband-IF TX and/or RX RF front-end. The IF-baseband transceiver interfaces with the RF front-end at an intermediate frequency of 3 GHz, which can be tuned anywhere between 0.2 and 7.8 GHz. Depending on the specific requirements of the experiment, an appropriate RF front-end is selected. For example, certain experiments, such as those involving monostatic sensing, require both the TX and RX RF front-ends to be co-located. In the following sections, we provide detailed descriptions of the IF-baseband transceiver and the RF front-end implementations.

\subsection{IF-Baseband Transceiver}
\label{sec:if-baseband}
\begin{figure}[!t]
\centering
\subfloat[]{%
  \includegraphics[clip,width=0.45\textwidth]{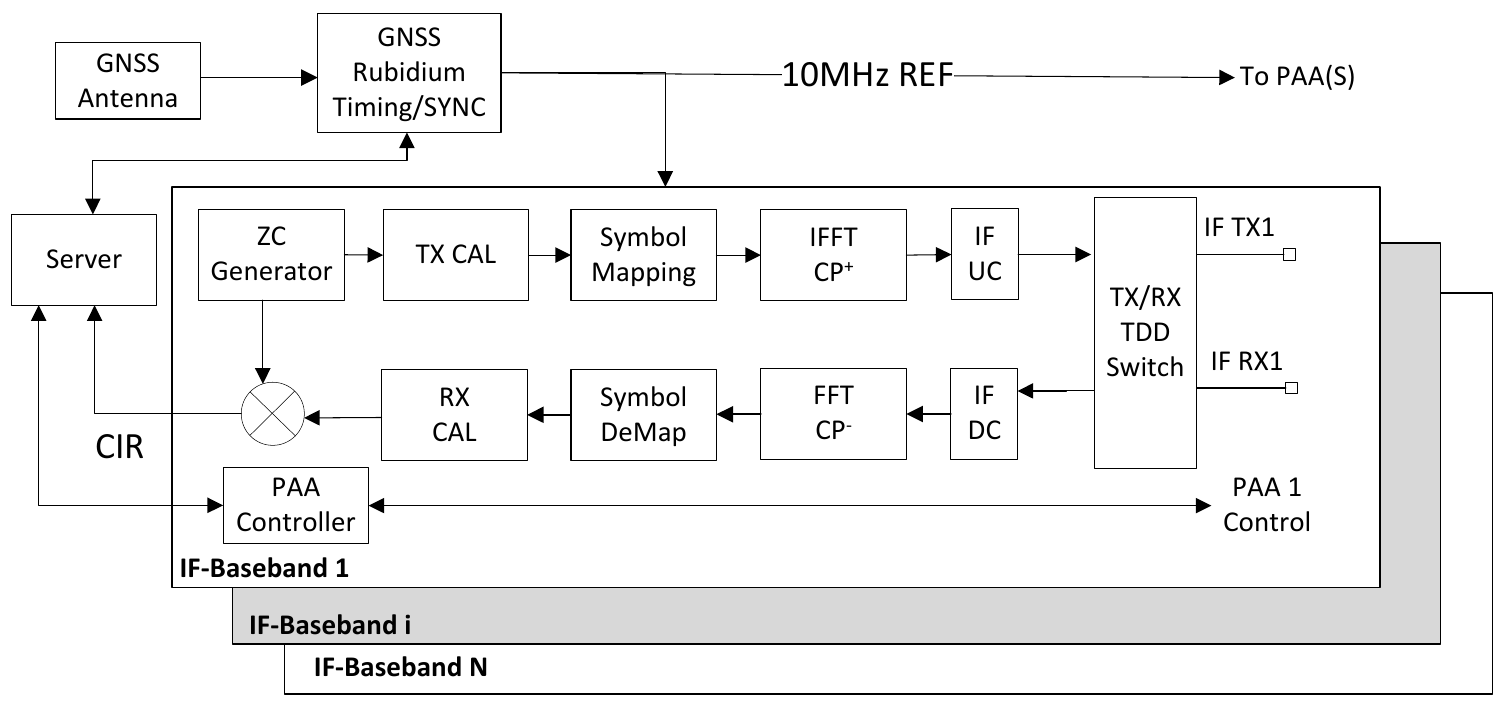}%
  \label{fig:If-BBfunc}
}

\subfloat[]{%
  \includegraphics[clip,width=0.45\textwidth]{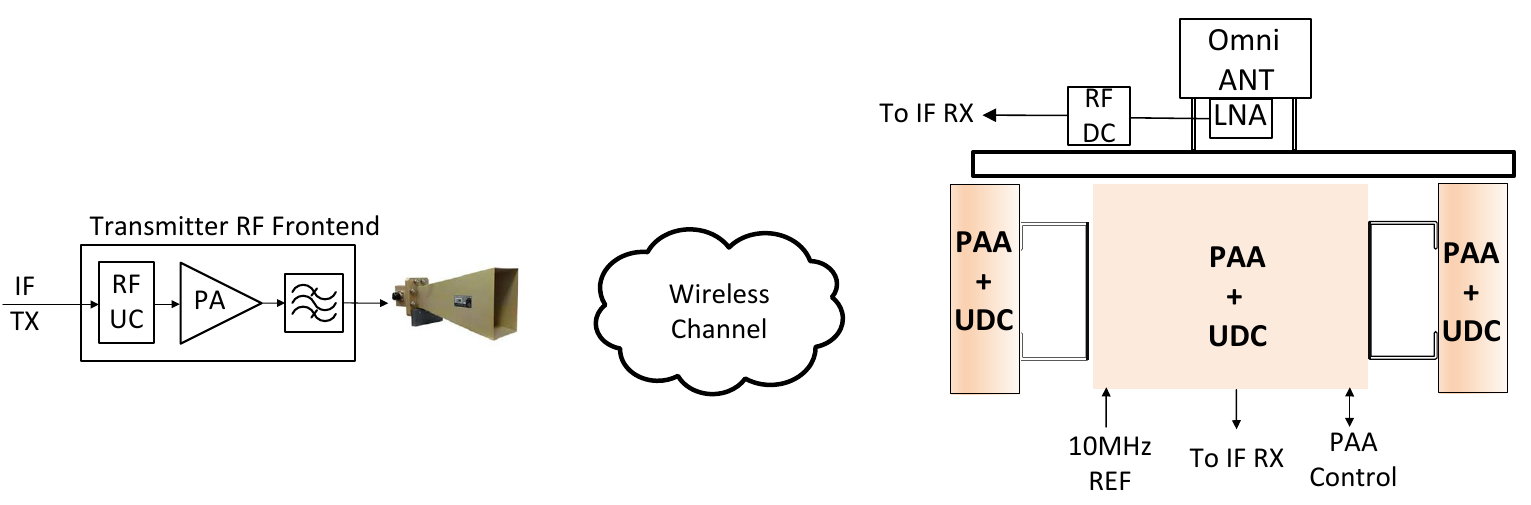}%
  \label{fig:RFfunc}
}

\caption{Functional block diagram of the channel sounder (a) IF-Baseband system, (b) RF front end used in this work}

\end{figure}

The IF-baseband transceiver is a modular piece of hardware and software that is the core to our channel sounder. As shown in Fig.~\ref{fig:If-BBfunc}, the IF-baseband transceiver consists of multiple software defined radios (SDRs), a timing/synchronization network, and an x64 server. Multiple TX and RX IF ports are supported and can be scaled as needed to meet the needs of a given experiment. An additional digital control interface is used to control phased array modules (when present) using synchronous digital commands. This is implemented in the SDRs' onboard FPGA general purpose input/output pins using logic driven off the data clock. This ensures phased array beam configuration and TX/RX samples are tightly synchronized to global system timing. 

In the TX signal path, a ZC sequence is mapped onto subcarriers. The ZC sequence is repeated four times, to get an averaging gain of 6 dB. The ZC sequence is then modulated via OFDM and upconverted inside the SDR FPGA to IF. In the RX signal path, the IF signal is downconverted to baseband and demodulated using a standard OFDM receiver. The baseband samples are correlated with a copy of the transmit ZC sequence to yield the complex channel impulse response (CIR). CIRs can be recorded at up to 25 GB/s sustained for high mobility and/or channel count experiments. Tables \ref{tab:transmitter} and  \ref{tab:receiver} provide the specifications for transmitter and receiver, respectively.

Multiple IF-Baseband systems are synchronized to each other using global timing reference and a transmission and reception schedule. This timing reference can be derived from multiple sources including GNSS, Rubidium oscillator, or precision time protocol (PTP), depending on the needs of the experiment and whether GNSS signals are available.

\begin{table}[t!]

	\caption{Transmitter Specification}
	\begin{center}
		\begin{tabular}{| l | c |}
			\hline
			\textbf{Attributes} & \textbf{Values}\\
			\hline

			Frequency (GHz) & 7, 8.3, 11.3, 14.5 \\
						\hline
			Bandwidth (MHz) & 400 \\
									\hline
			Modulation & OFDM \\
									\hline
			Subcarrier spacing (kHz) & 120 \\
									\hline
			ZC Sequence Length & 3343\\
									\hline
			Antenna, Gain (dBi) & Standard horn, 10  \\
									\hline
			Antenna 3-dB beamwidth & AZ: 55$^{\circ}$, EL: 55$^{\circ}$ \\
									\hline
			Polarization & Vertical\\
									\hline
			Maximum EIRP (dBm) & +43\\
									\hline
			
			\hline
		\end{tabular}
		\label{tab:transmitter}
	\end{center}
\end{table}
\begin{table}[t!]
		\caption{Receiver Specification}
		\begin{center}
			\begin{tabular}{| l | c |c |}
				\hline
				\textbf{Attributes} & \multicolumn{2}{|c|}{\textbf{Values}} \\
				\hline

				Antenna System& Omnidrectional & Phased Array \\
								\hline
				Frequency (GHz) & 7,8.3,11.3,14.5 & 8.3,11.3,14.5 \\
								\hline
				Min. MPC Delay Resolution (ns) & \multicolumn{2}{|c|}{2.5} \\
								\hline
				No. of antenna elements & (2) Omni\footnotemark[1] & 32 or 64\footnotemark[2]\\
								\hline
				No. of beams/array (8/11/15 GHz) & N/A &  15/15/20 \\
								\hline
                Field of view &  \multicolumn{2}{|c|}{AZ: {360\textdegree}, EL:{$\pm$32.5\textdegree}} \\
                				\hline
				Receiver type & \multicolumn{2}{|c|}{Full correlator}\\
								\hline
                    Processing Gain (dB) & \multicolumn{2}{|c|}{41}\\
								\hline
				Number of receivers & 1 & 4 (1 per array) \\
								\hline
				360\textdegree\  CIR acquisition time & $<$ 40 $\mu$s& 0.5-0.9 ms\\
								\hline
				Number of averaging & \multicolumn{2}{|c|}{User selectable} \\
								\hline
				Max. Meas. Excess Delay& \multicolumn{2}{|c|}{8 $\mu$s} \\   		
						\hline
				Receiver noise figure& 1.5 dB & 8.3 dB \\
				\hline
                \multicolumn{3}{l}{\scriptsize$^1$Two Omni antennas may be used with any phase array frequency.}\\
                \multicolumn{3}{l}{\scriptsize$^2$32 elements for 8.3, 11.3 GHz arrays and 64 for 14.5 GHz array.}\\
			\end{tabular}
    			\label{tab:receiver}
		\end{center}
 
	\end{table}

\subsection{The RF Front-End Transceiver}
\label{sec:rf}

The TX RF front-end upconverts the IF signal using a banded upconverter module. The RF signal is then amplified using a 10 W wideband power amplifier and transmitted through a standard horn antenna. After calibration, the radiated signal is frequency flat with less than 1 dB ripple and accurate to within 1 dBm EIRP.

The experiments described in this paper used a horn antenna for RF signal transmission, however double-directional channel sounding can be achieved by replacing the entire RF front-end with a phased antenna array module.

\begin{figure}[!b]
		\centerline{\includegraphics[width=\columnwidth]{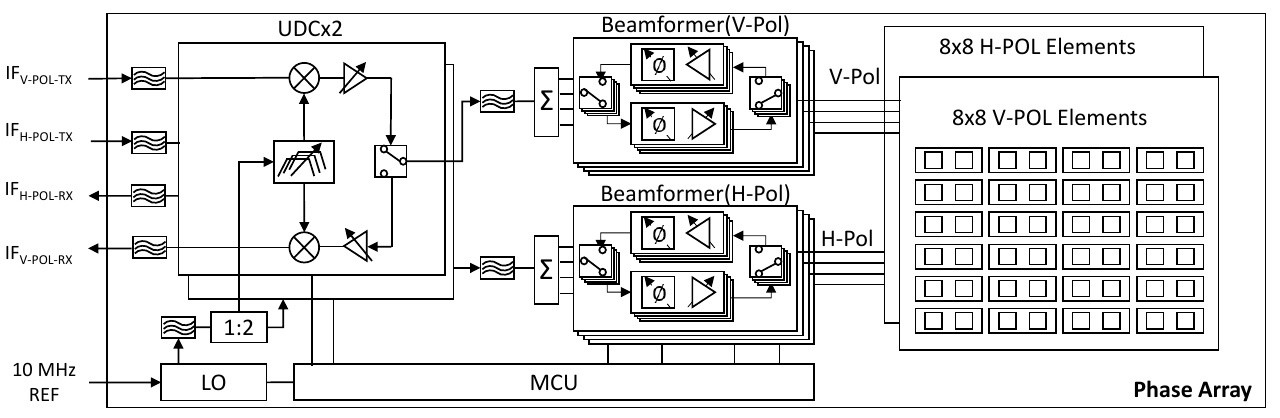}}
		\caption{Functional block diagram of the phase array module.}
		\label{fig:RF-frontend}
	\end{figure}

The RX RF front-end is comprised of an omni antenna subsystem and a phase array subsystem, as shown in Fig. \ref{fig:RF-frontend}. The omni subsystem comprises of an antenna, low noise amplifier, and downconverter. The phase array subsystem includes the beamformers, up/down converter (UDC) unit and a micro controller unit (MCU) for controlling the beam and phase array. The antenna is well simulated and calibrated in practice to achieve less than 6 dB scan loss with beam scan range of ±60{\textdegree} and ±45{\textdegree} in azimuth and elevation, respectively. Custom beam pointing angles and beam shapes were achieved by utilizing the phase shifter and variable gain amplifier inside the beamformer which is controlled by the MCU. The UDC converts the sounding signal from IF/RF to RF/IF with a dedicated RF filter to eliminate the unwanted signal and out of band emissions. The array's local oscillator module synthesizer is synchronized to the baseband unit by an external reference clock.
	
\section{Calibration}
\label{sec:cal}

Calibration of a channel sounder is an essential step to ensure accurate and reliable measurements of the wireless channel parameters. Each day, before and after the measurement campaign, the channel sounder is calibrated using a known reference signal. This process ensures that measured channel characteristics are not affected by any biases or errors introduced by the measurement equipment. We require four steps to calibrate our channel sounder: 

\begin{figure}[t!]
	\centering
	\subfloat[OTA calibration of the channel sounder at reference distance.]{
		\includegraphics[clip,width=0.45\textwidth]{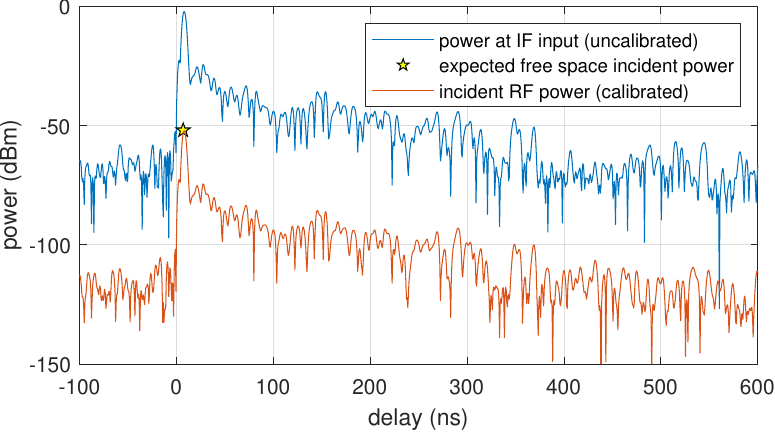}
		\label{fig:cal1}
	}
	
	\subfloat[PDPs of omnidirectional antenna vs sum of phased array antenna beams.]{
		\includegraphics[clip,width=0.45\textwidth]{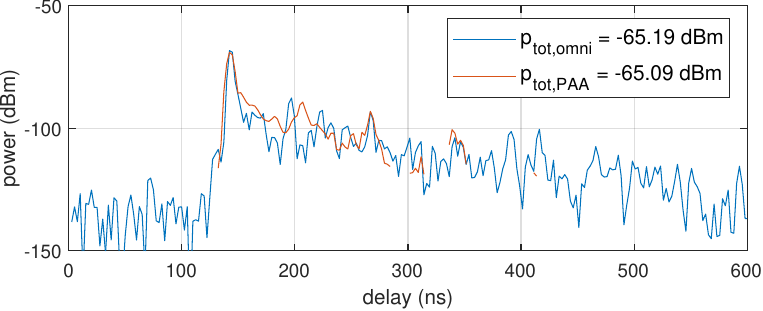}
		\label{fig:cal2}
	}
	
	\caption{OTA sounder calibration}
			\label{fig:cal}
\end{figure}

\subsubsection{Transmit Flatness and EIRP} The transmitter is calibrated at the transmit antenna input port(s) using a spectrum analyzer at each operating band. Target transmit power is the target EIRP minus the gain of the transmit antenna. A vector of coefficients is computed per transmit IF port and multiplied to the IF signal. The transmit radiated power is calibrated to within 1 dBm and 1 dB ripple over the transmission band. 
    \subsubsection{Receive Flatness} If the RX response is not sufficiently flat, the RX is calibrated for frequency flatness at each IF receive port. The frequency flat TX from Step 1 is connected to each RX port using a cabled network with known frequency response. Similar to Step 1, the calibration coefficients are multiplied across frequency to each receive IF port.
    
    \subsubsection{Receive Incident Power} Once the transmitter and receiver are frequency flat, the system is calibrated for incident RF power over the air in the field. This includes the antenna response in the calibration as well as any additional power mismatches that may occur during fielding of the system. For this step, the transmitter and receiver are placed in a location with known separation and good line of sight visibility. The receiver is calibrated to the expected RF incident power at the antenna for each band using the free-space channel tap. A power delay profile (PDP) is defined as the magnitude squared of the CIR. In Fig. \ref{fig:cal1} we show a typical PDP at a calibration reference TR-separation distance of 3 m before and after calibration. It is seen that the incident RF power matches the free space received power. This calibration process is repeated four times by freely rotating the phased array antenna platform to align with the peak gain directions. 
    \subsubsection{Omnidirectional Power from Phased Array Beams}  The omnidrectional PDP is synthesized from the PDP of each beam by first thresholding noise power per PDP per beam, followed by summing PDPs across all beams. The total omnidirectional power is computed from the omnidirectional PDP by linearly summing the power contained in all the channel taps of the thresholded omnidirectional PDP. Fig.~\ref{fig:cal2} shows the sum power of beams.   Total power and profile should match that of the closely co-located omnidirectional antenna. In this case, total receive power differs by a mere 0.10 dBm.

\section{Channel Sounder Performance}
\label{sec:field_test}
\begin{figure}[t!]
	\centering
	\subfloat[Normalized power vs. time delay]{
		\includegraphics[clip,width=0.45\textwidth]{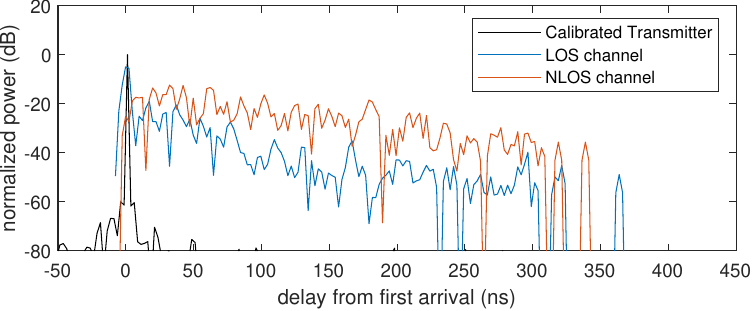}
		\label{fig:cir}
	}
    
	\subfloat[Normalized power vs. frequency]{
		\includegraphics[width=0.45\textwidth]{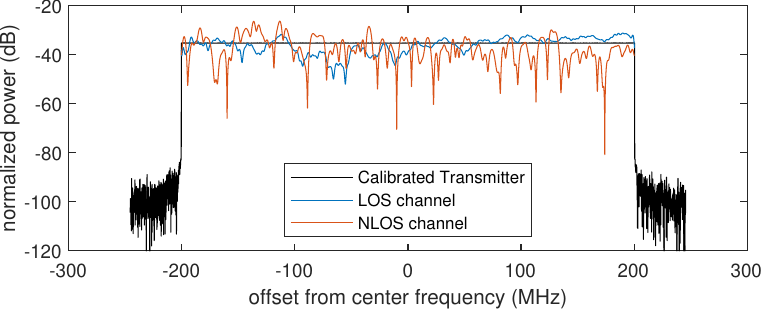}
		\label{fig:cfr}
	}
	\caption{Normalized received power vs. delay (a) and frequency (b) in LOS and NLOS environments as compared with the normalized TX power.}
			\label{fig:cifr}
\end{figure}

Experiments were performed to verify the channel sounder's measurement capability, their validity and repeatability under certain controlled conditions. Fig. \ref{fig:cifr} shows exemplary power distribution over delay and frequency. Power is normalized to total power and is shown as measured at the TX and at the RX in LOS and NLOS environments. Measurements were performed in 7, 8.3, 11.3 and 14.5 GHz spectrum bands \cite{fcc_license}. Next, we present some of our outdoor results taken at 7 and 14.5 GHz during these experiments.

\subsection{Experiment Procedure}
 The performance verifications test were performed within a 0.5 mile-long outdoor testbed. We chose this outdoor location because of the nature of LOS/NLOS channel conditions with little to no vehicle traffic, known location of clutters, etc. In this paper, we show the performance of the sounder when the height of TX antenna is set to 10 m above ground level. The receiver was then moved along a predetermined 300-meter-long outdoor path. At each location, CIR measurements were made to form a dataset over a range of distances. This allowed us to evaluate the performance of the sounder over long distances as well as over extended time periods.

\subsection{Key Findings}
There are many metrics to fully characterize the performance of the channel sounder. Here are some noteworthy findings:  
\subsubsection{Received Power and Path Loss}

\begin{figure}[!t]
\centering
    \includegraphics[clip,width=\columnwidth]{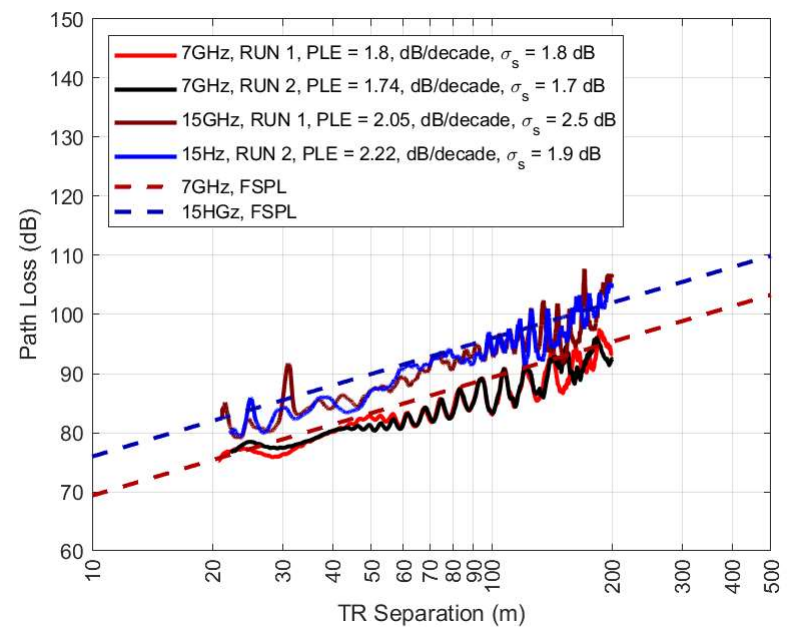}%
    \caption{The scatter plot of the path loss vs T-R separation at 7 and 14.5 GHz.}
        \label{fig:Outdoor_LOS}
\end{figure}

Fig. \ref{fig:Outdoor_LOS} shows the scatter plot of path loss as a function of TR separation for two separate experiments over the same path but two different frequency bands. A least square linear regression line is fit through the measured path loss points to minimize the mean square error. As seen in Fig. \ref{fig:Outdoor_LOS}, the regression line has a slope of ten times the path loss exponent (PLE) and a standard deviation of $\sigma_S$, called shadow fading (due to randomness of different level of clutter at the same distance).  We compared the results of our findings with that of Free Space Path Loss (FSPL) and look at the Power-Angular Profile (PAP) and Power-Angular Delay Profile (PADP) at a specific location within this database. The repeatability of the sounder is verified by conducting two separate runs over the same path. The $PLE$ and $\sigma_S$ stay withing tenths of dB over the two runs. 

\subsubsection{PAP and PADP}
\begin{figure}[t!]
	\centering
	\subfloat[2D Omnidirectional PAP.]{
		\includegraphics[clip,width=0.45\textwidth]{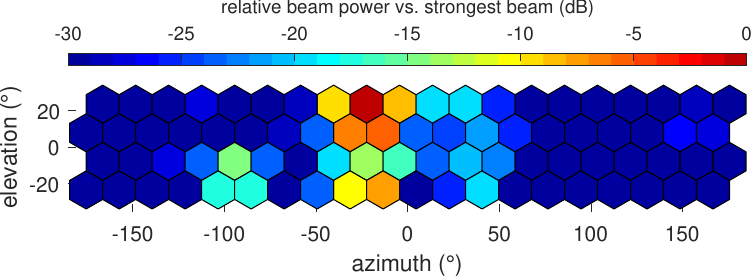}
		\label{fig:angular1}
	}
	
	\subfloat[Omnidirectional Azimuth PADP]{
		\includegraphics[clip,width=0.45\textwidth]{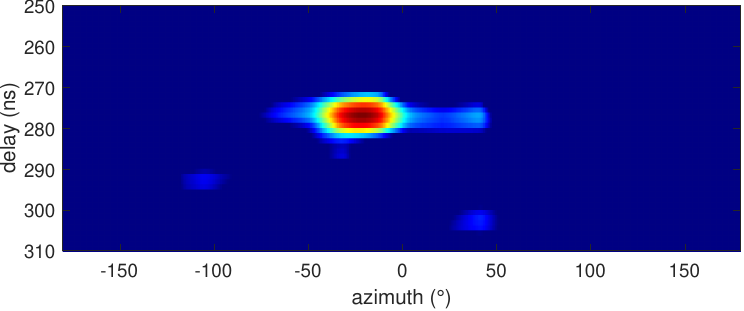}
		\label{fig:angular2}
	}
    
	\subfloat[Omnidirectional Elevation PADP]{
		\includegraphics[clip,width=0.45\textwidth]{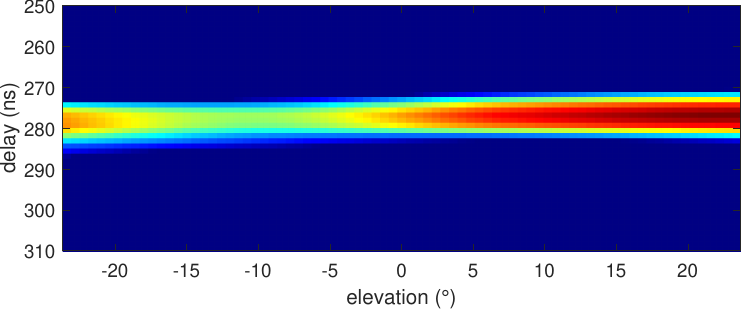}
		\label{fig:angular3}
	}
	\caption{Measured Outdoor Power-Angular Profile LOS channel at d = 76 m at $f_c$ = 14.5 GHz.  }
			\label{fig:angular}
\end{figure}

Fig. \ref{fig:angular1} shows the PAP and Figs. \ref{fig:angular2} and \ref{fig:angular3} show the PADP of a LOS channel in the azimuth and elevation plane at 14.5 GHz. The RX was 76 m away and {20\textdegree} off the boresight axis of the TX antenna. As seen in Fig. \ref{fig:angular2}, the the sounder detects strong multipath components between azimuth angles $-50^\circ$ and $0^\circ$. As is visible in Figs. \ref{fig:angular1} and \ref{fig:angular3}, the multipath are arriving from an elevation angle of $20^\circ$. We can also note multiple weaker ground and tree reflections as captured by the bottom row of beams in Fig. \ref{fig:angular1}.

\subsubsection{ISAC}
\begin{figure}[t!]
	\centering
		\includegraphics[width=0.40\columnwidth]{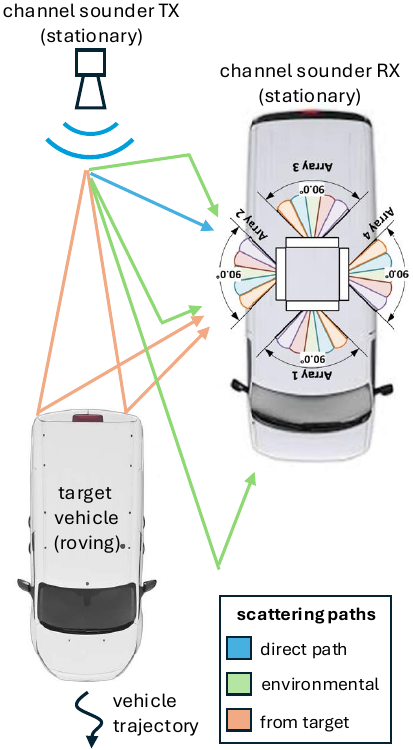}
		\includegraphics[width=0.5\columnwidth]{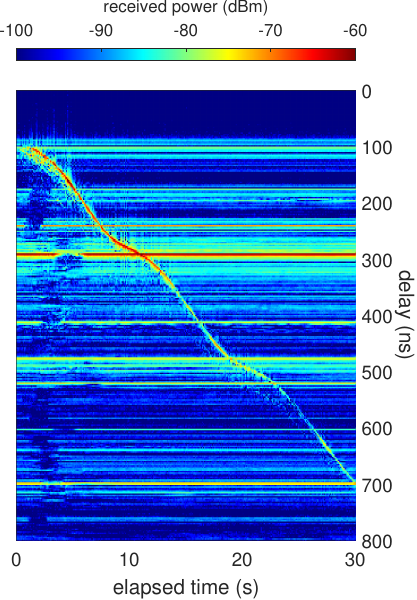}

	\caption{7 GHz bi-static sensing scenario (left) and measured PDPs of channel as target vehicle drives away (right).}
			\label{fig:isac}
\end{figure}

The evaluation of ISAC channels is a key requirement of our channel sounder. Fig.~\ref{fig:isac} shows an example of one such scenario where we emulate bi-static sensing on the downlink in an outdoor microcell. The channel sounder TX is mounted at 10 m height in a cul-de-sac. The RX is mounted on the rooftop of a van 25 m away. A target vehicle is driven from close proximity of RX to a maximum distance of $\sim$200~m. TX and RX remain stationary during the duration of the experiment. Fig.~\ref{fig:isac} (right) shows the measured PDP of the channel over 30~s. A significant portion of the PDP remains static due to static environmental scattering. The scattered energy from the target can be seen in the diagonal line starting at at 0 s elapsed time at $\sim$100~ns delay. The received scattered power from the target drops as the vehicle drives 200~m down range ($\sim$700 ns bi-static delay). The environmental scatterers are disturbed due to blockage by the target vehicle in the 2-8~s portion of the experiment but are re-established as soon as the vehicle passes by. 

\begin{figure}[t!]
	\centering
		\includegraphics[width=\columnwidth]{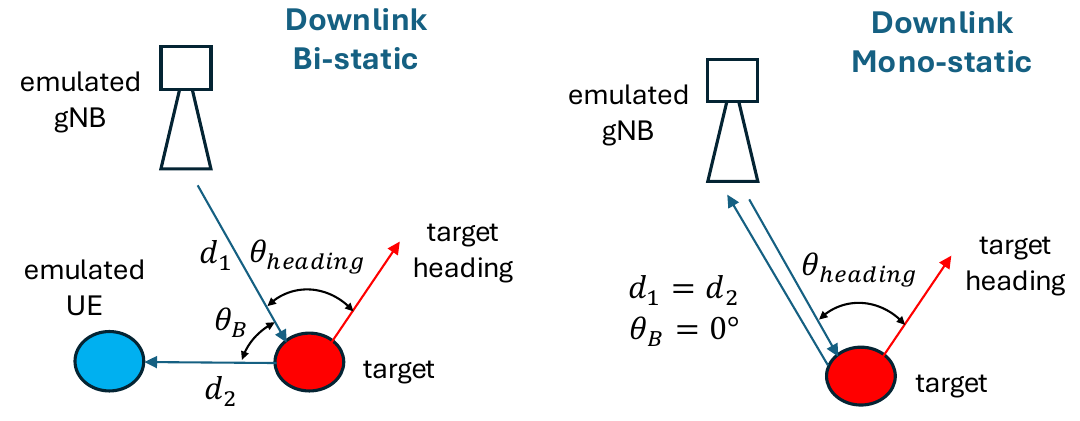}
	\caption{Sensing scenarios measured and modeled.}
			\label{fig:isac_scenarios}
\end{figure}

To obtain a more comprehensive view of ISAC channels, we configured our system for both mono-static and bi-static sensing. Additional experiments were conducted with different TX/RX configurations to develop a model for the large scale ISAC sensing channel. Fig.~\ref{fig:isac_scenarios} shows the downlink bi-static and mono-static configurations for sensing. Here, downlink transmission from a 5G gNB is emulated using the channel sounder TX. which illuminates a mobile target. The target has a 2D heading angle $\theta_{heading}$. Scattered power from the target is collected at either the TX location (emulating a mono-static gNB) or at the van-mounted receiver shown in Fig.~\ref{fig:isac}. Two different targets were measured--(\textit{i}) the vehicle target shown in Fig.~\ref{fig:isac} and (\textit{ii}) a walking pedestrian. In both cases, a GNSS receiver was attached to the target so that its precise location was known during the experiment. The delay of the target scatterer was calculated from the known location of TX and RX. The scattered energy from the target was isolated from the environment and analyzed independently.  

We developed $PL_t$, a target path loss model, for ISAC where $PL_t$ can be expressed as follows: 
 \begin{align}
 \label{eq:PL}
     PL_t &= PL_1 - G_s + PL_2 \:\mathrm{[dB]},\nonumber
 \end{align}
where $G_s = \gamma + 10\log_{10}\frac{4\pi}{\lambda^2}\:\mathrm{[dB]}$ is the scattering gain of incident RF power and $\gamma$ is the radar cross section (RCS) of the target. $PL_1$ and $PL_2$ are the path loss associated to path lengths $d_1$ and $d_2$ (See Fig.~\ref{fig:isac_scenarios}) and computed from FSPL as $20\log_{10}(d_1f_c)+32.44-G_t$ and $20\log_{10}(d_2f_c)+32.44-G_r$, respectively. $f_c$ is the center frequency, and $G_t$ and $G_r$ are the TX and RX antenna gains, respectively. We model $\gamma$ as a log-normal random variable ($\log \gamma \sim \mathcal{N(\mu,\sigma)}$) with no dependence on $\theta_B$ or $\theta_{heading}$. Table~\ref{tab:rcs} shows the model parameters for the RCS of the two target classes measured. As expected, due to the smaller size of the pedestrian, it has lower scattering power versus that of the vehicular target. Additionally, we confirm that the mono-static RCS of a target is significantly higher than the bi-static RCS, a property that is common for most target geometries~\cite{Skolnik_2008}.

\begin{table}[t]
	\caption{RCS parameters for two target classes.}
	\begin{center}
		\begin{tabular}{| c | c | c | c | c |}
			\hline
\textbf{Target} & \multicolumn{2}{|c|}{\textbf{bi-static}}& \multicolumn{2}{|c|}{\textbf{mono-static}}\\ \cline{2-5}
\textbf{Class}  & \textbf{$\mu$ (dBsm)} & $\sigma$ (dBsm) & \textbf{$\mu$ (dBsm)} & $\sigma$ (dBsm) \\
				\hline
                Passenger car & -0.1 & 6.1 & 7.7 & 8.4 \\
                \hline
                Pedestrian & -14.4 & 6.7 & -6.2 & 10\\
                        
			\hline
		\end{tabular}
		\label{tab:rcs}
	\end{center}
\vspace{-8pt}
\end{table}

\section{Conclusion}
\label{sec:conclusion}

The advanced channel sounding system described in this paper is a highly adaptable and precise tool for measuring and characterizing omni-directional wireless channel parameters. Its notable features include high resolution, sensitivity, and a rapid measurement capability of up to 80 beams in a 360-degree sweep in under 0.9 milliseconds. With a 400 MHz instantaneous bandwidth and a multipath resolution of 2.5 nanoseconds, the system reliably measures path loss up to 170 dB. It can be easily reconfigured for different frequency bands by changing the phased array antennas. This versatile system is suitable for a wide range of applications, including indoor and outdoor measurement campaigns, doubly directional channel sounding, high-speed vehicular communication experiments (such as V2V and V2I), and integrated communication and sensing experiments. These attributes make the system an essential tool for advancing research and practical applications in wireless communication systems, ultimately contributing to the development of more efficient and reliable wireless technologies.

\bibliographystyle{IEEEtran}
\bibliography{references}

\end{document}